\documentstyle[11pt,newpasp,twoside,epsf]{article}
\def\edcomment#1{\iffalse\marginpar{\raggedright\sl#1\/}\else\relax\fi}
\reversemarginpar

\def\@journalname{ASP Conference Series}
\def\cpr@holder{Astronomical Society of the Pacific}
\def\@jourvol{??}
\def\cpr@year{2000}
\def\vol@title{Massive Stellar Clusters}
\def\vol@author{A.\,Lan\c{c}on and C.M.\,Boily, eds.}

\newcommand{\Msun}{\mbox{${\cal M}_\odot$}}
\newcommand{\tcross}{\mbox{$t_{\rm cross}$}}
\newcommand{\etal}{et al.}
\newcommand{\kms}{\mbox{km\thinspace s$^{-1}$}}

\begin{document}
\title{Star Clusters and the Duration of Starbursts}
 \author{G.R.\ Meurer}
\affil{Department of Physics and Astronomy, The Johns Hopkins
University, Baltimore, MD 21218, USA}

\begin{abstract}
The duration of starbursts is important for determining how they are
regulated and the impact they have on their environment.  Starbursts
contain numerous prominent star clusters which typically comprise in
total $\sim$ 20\%\ of the ultraviolet light, embedded in a diffuse glow
of recently formed stars responsible for the dominant $\sim 80$\%.
Hubble Space Telescope images have been obtained for four starburst
galaxies in order to determine their burst duration from the ages of
their star clusters.  Preliminary results for NGC~3310 are reported
here.  The $UBI$ colors of its clusters and diffuse light give
consistent results.  The clusters have a broad range of colors,
consistent with a population of instantaneous bursts with ages ranging
from 0 to a few 100 Myr.  The diffuse light has a narrow range of color
consistent with continuous star formation over timescales ranging from
$\sim 10$ to 100 Myr.  Hence NGC~3310's starburst has lasted $\sim 100$
Myr, or about 10 times the crossing time.  Other results noted in the
literature also indicate bursts lasting several times longer than the
crossing time.  These results suggest that starbursts are not
self-extinguishing flashes.  Rather they are sustainable, perhaps
self-regulated.
\end{abstract}


\section{Introduction}

Over the past few decades starbursts, brief intense episodes of massive
star formation, have been recognized as important agents of galaxy
evolution.  However, despite many journal pages per year of attention,
there is much we do not know about the inner workings of starbursts.  In
particular: how long do they last?  Estimates in the literature have
been obtained from a variety of techniques and range from Myr
to Gyr time scales (Mas-Hesse \&\ Kunth \&\ 1999, and Coziol,
Barth \&\ Demers 1995).  {\em A priori\/} the minimum expected duration
is the crossing time \tcross: the time it would take for a disturbance
to travel from one end of the starburst to the other.  Local starbursts
typically have size scales of $\sim 0.2 - 2$ Kpc and velocity
dispersions of $\sim 30 - 300$ km/s, and hence crossing times ranging
from $\sim 1$ Myr to $60$ Myr, with 10 Myr being typical.

Starburst duration is important for determining the efficiency that
metals are released into the IGM, and for determining the total number
of bursts that a galaxy can undergo.  Knowing the duration of
starbursts will give us a better understanding about how bursts
evolve.  If the durations are similar to \tcross\ it would suggest that
they are self-extinguishing explosions, destroyed by their energy output
produced in a non-equilibrium fashion.  Conversely, durations
much longer than \tcross\ would indicate that starbursts are
sustainable, perhaps even being in equilibrium: i.e.\ ISM inflow
$\approx$ star formation rate (SFR) + outflow.

\section{Starbursts and star clusters.}

In the mid-1990's we obtained HST ultraviolet (UV) images of starburst
galaxies with the {\em Faint Object Camera\/} imaging at $\lambda
\approx 2300$\AA\ in order to examine the distribution of high mass
stars which power starbursts.  The results of our imaging study are
given in Meurer \etal\ (1995; hereafter M95).  The UV structure of
starbursts is well exemplified by NGC~3310 and NGC~3690 as shown in
Fig.~1.  Immediately striking are numerous prominent and compact star
clusters.  However, the total UV emission is dominated by a diffuse
distribution of stars.  Typically, this amounts to about 80\%\ of the UV
light within starbursts (M95).  The nearest starbursts in our sample
start to resolve into individual high mass stars which trace the diffuse
light isophotes, demonstrating that the diffuse light is not a product
of scattered light from the clusters, nor is it a figment of the
pre-COSTAR optics of HST.  These images reveal that a starburst is not
the same thing as a star cluster.  Nor, is it the sum of multiple
clusters.  The natural conclusion is that there are two modes of star
formation in starbursts: a very significant 20\%\ of the star formation
is in compact star clusters while the dominant mode is diffuse star
formation.

\begin{figure}
\plottwo{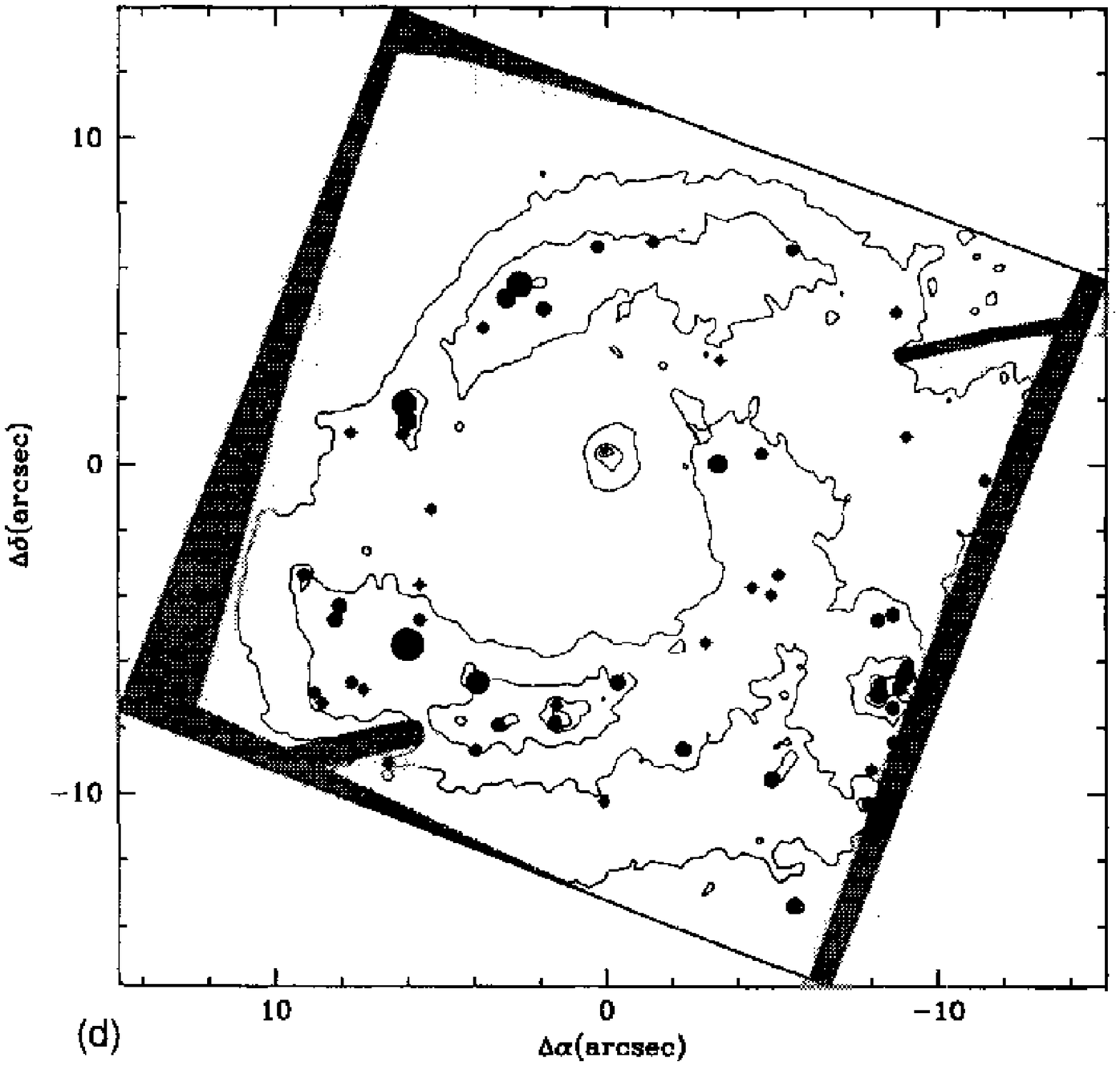}{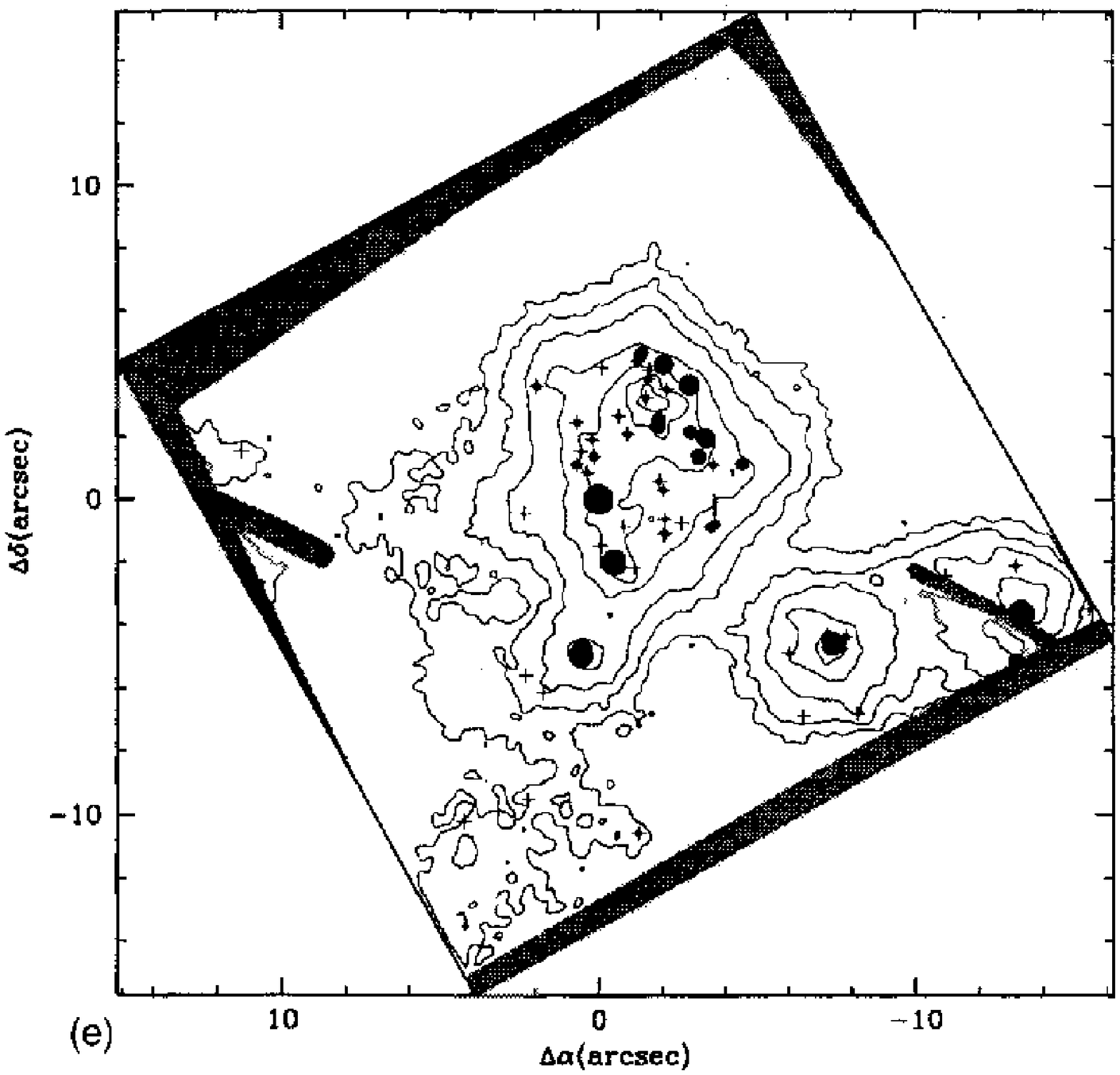}
\caption{Schematic structure of the starburst galaxies NGC~3310 (left)
and NGC~3690 (right) as seen in the ultraviolet (2300\AA) by the {\em
Faint Object Camera} on HST (M95).  The dots and crosses mark star
clusters, with size indicating brightness.  Isophotes show the
distribution of diffuse light (after all clusters and their PSF halos
are subtracted).  The images have been rectified and oriented with north
up, and east on the left.  The distorted edges and occulting fingers of
FOC are outlined.  The two arcsec tick marks on the axis correspond to
175 pc and 430 pc for NGC~3310 and NGC~3690
respectively. }
\end{figure}

The $\sim 10$ Myr crossing time of starbursts is set by the dimensions
of the diffuse light.  It is difficult to verify whether the star
formation timescales are consistent with such large crossing times.
Broad band photometric indexes are degenerate in terms of population age
$t$ and duration $dt$ (Mas-Hesse \&\ Kunth 1991).  This is because for
extended duration star formation it is the youngest stellar populations
that dominate the bolometric output.  Hence the light from the onset of
the burst has relatively little weight.

The clusters within starbursts are very compact, having effective radii
typically of a few pc (M95, Whitmore \etal\ 1999).  They have velocity
dispersions of typically $\sim 10$ \kms (Ho \&\ Fillipenko, 1996; Smith
\&\ Gallagher, 2000), and hence crossing times less than a Myr.  Since
their photometric ages are usually at least a few Myr, they are well
mixed systems.  It is fair to assume that they were each formed in a
single short duration burst.  If so, one can determine good photometric
ages; since $dt$ is very small the $t - dt$ degeneracy is broken.

This is the basis of our project to measure the the duration of
starbursts as a whole using star clusters as chronometers.  The duration
of the entire starburst is the width of the cluster $t$ distribution.
We have been granted HST time in cycle 6 to measure the starburst
duration in four systems.  The sample is listed in Table 1, which lists
the distance ($D$), effective radius of the entire starburst ($R_e$),
velocity dispersion, $\sigma$, and crossing time ($t_{\rm cross} =
2R_e/\sigma$).

\begin{table}
\caption{WFPC2 sample}
\begin{tabular}{l r r r r} 
&&&&\\[-2mm] \tableline &&&&\\[-2mm]
Galaxy      &  $D$  & $R_e$ & $\sigma$ & $t_{\rm cross}$ \\
            & [Mpc] &  [pc] &  [km/s]  &   [Myr]   \\ 
&&&&\\[-2mm] \tableline &&&&\\[-2mm]
NGC~4670    &   15  &  220  &     42   &     10    \\
NGC~3310    &   18  &  680  &    128   &     10    \\
Tol1924-416 &   37  &  480  &     38   &     25    \\
NGC~3690    &   44  &  680  &    306   &      4    \\
&&&&\\[-2mm] \tableline\tableline
\end{tabular}
\end{table}

These galaxies were imaged with WFPC2 using the F336W ($U$), F435W
($B$), and F814W ($I$) filters.  While our own data on Tol1924-416 has
not yet been obtained, \"Ostlin \etal\ (1998) have obtained equivalent
data for this system, known in Europe as ESO 338-IG04.  Three color
images of these systems can be found on the web \footnote{\sf
http://www.pha.jhu.edu/$\sim$meurer/research/starburst\_dt.html}.  Most
of the rest of this talk will focus on NGC~3310 for which the analysis
has progressed the furthest.

\section{NGC~3310}

NGC~3310 is a nearly face-on spiral with a prominent ring of star
formation encircling its mildly active nucleus (LINER or transition
spectrum: Heckman \&\ Balick 1980; Pastoriza \etal\ 1993).  In ground
based images the ring is very clumpy, with the brightest clump described
as a ``Jumbo'' H{\sc ii} region by Balick and Heckman (1981).
NGC~3310's outer structure is disturbed (van der Kruit \&\ de Bruyn,
1976) suggesting that an interaction or merger has triggered the present
starburst.  The spectroscopy of Pastoriza \etal\ (1993) reveals both the
Wolf-Rayet feature at 4686\AA\ and the Ca{\sc ii} triplet, which arises
in red supergiants, indicating that star formation has proceeded for
$\ga$ 10 Myr.

The $U$ band Planetary Camera chip image of NGC~3310's center is shown
in Fig.~2.  The structures detected from the ground are clearly revealed
including the starburst ring and the Jumbo H{\sc ii} region.  Numerous
compact clusters are prominent in these structures.  NGC~3310 is close
enough, and the images deep enough that some of the faintest sources are
probably individual high-mass stars.

\begin{figure}
\plottwo{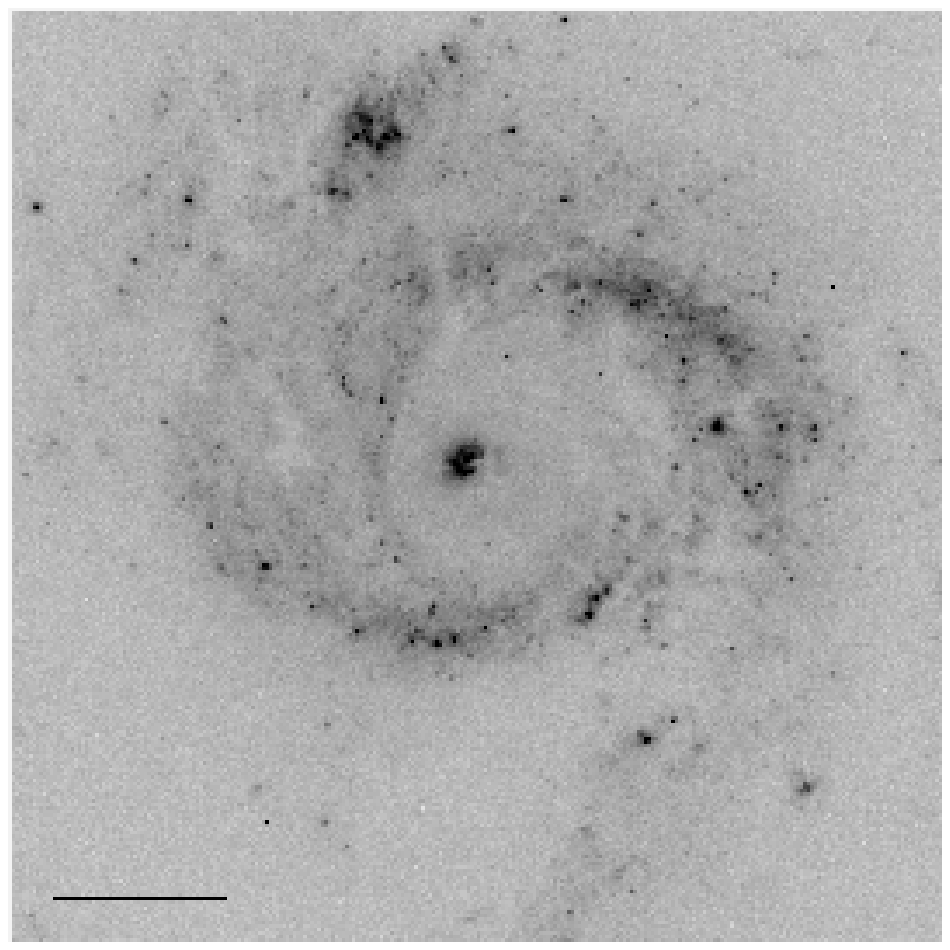}{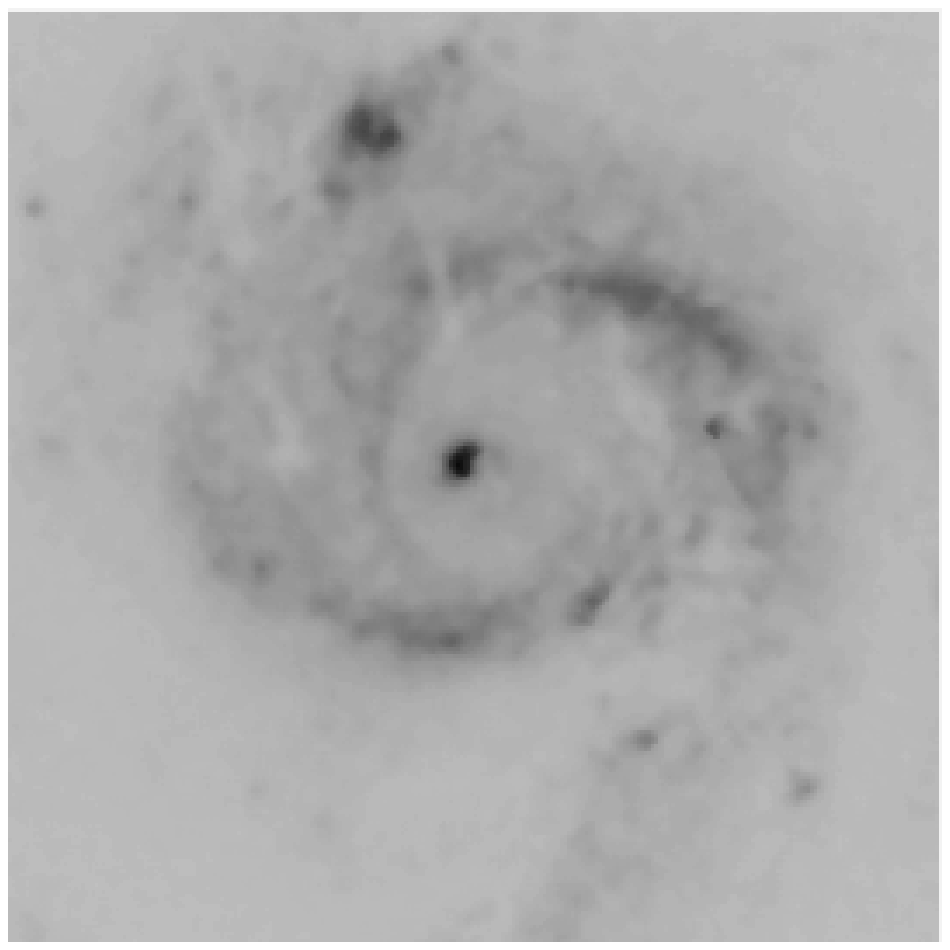}
\caption{$U$ band (F336W) PC image of the nuclear starburst ring of
NGC~3310 is shown at left.  The scale bar is 500 pc long.  Note the
numerous prominent star clusters.  The diffuse light only image is shown
in the left hand panel.}
\end{figure}

Photometry was done on the clusters using DAOPHOT.  Isolated clusters on
the frame were used to construct the ``point spread function''.  So, it
is better described as a ``cluster spread function'', since the clusters
are comparable in size to the pixels, noticeably broadening the PSF
width (Whitmore \etal\ 1999; M95).  The clusters were separated from the
diffuse light and the photometry was refined in an iterative fashion.
The resulting total and diffuse light images are shown in Fig.~2.  A
comparison of the two panels clearly shows the dominance of the diffuse
light.  The fractional flux of NGC~3310 contributed by the clusters in
the different bands is 0.07 (UV; M95), 0.092 ($U$), 0.073 ($B$), and
0.065 ($I$).  Despite the improved optical performance of WFPC2 compared
to pre-COSTAR FOC, starbursts do not resolve out into an ensemble of
star clusters.

The clusters span a wide range of colors, as is apparent in the color
composite image presented at the meeting (see footnote 2).  While dust
lanes are apparent in the image, the lack of correlation between cluster
color and proximity to the dust lanes shows that dust is not causing the
effect.  This can also be seen in the $U-B$ versus $B-I$ two-color
diagrams shown in Fig.~3 where the clusters form an $S$ like locus; they
are not all strung out on a reddening vector.

\begin{figure}
\plotone{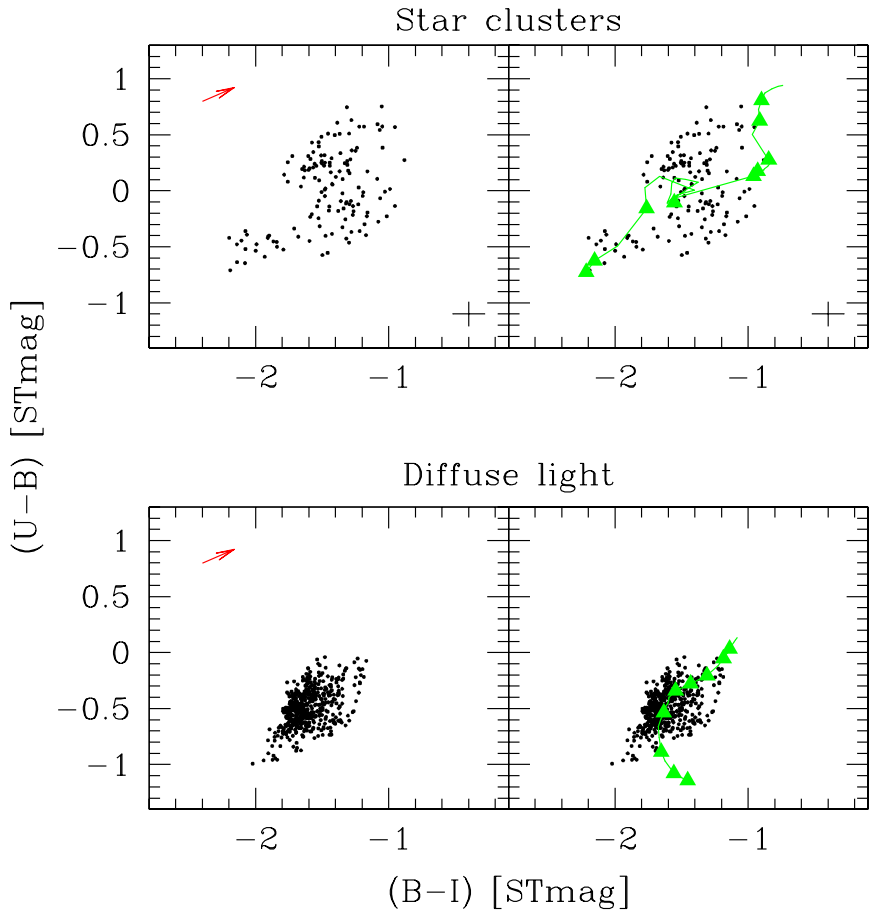}
\caption{Two color diagrams for the star clusters (top) and diffuse
light (bottom) in NGC~3310.  Only clusters with color uncertainties $<
0.15$ are shown.  The diffuse light points are for $\sim {\rm 80\, pc
\times 80\, pc}$ boxes excluding the nuclear region.  Comparisons with
Starburst99 population models (Leitherer \etal\ 1999) are shown in the
right hand panels. Instantaneous burst models are shown compared to the
clusters, while constant star formation rate models are compared with
the diffuse light.  The time/duration runs from 1 to 900 Myr with
triangles plotted at 1, 3, 5, 10, 30, 50, 100, 300, and 500 Myrs. A
Salpeter IMF over the mass range 1 to 100 \Msun, and SMC metallicity are
assumed and the models have been reddened by $E(B-V) = 0.1$ (arrows at
left).}
\end{figure}

In Fig.~3 the cluster data are compared with Starburst99 (Leitherer
\etal\ 1999) single burst models.  A range of metallicities and IMF
parameters were examined.  While none exactly matches the cluster two
color diagram (better models are needed), the data
matches the general sense of how photometric evolution proceeds.
Clusters start out at the lower left (blue in each color) and evolve
more rapidly in $B-I$ than $U-B$ as the main sequence widens and red
evolved stars appear.  After about 30 Myr, evolution becomes more rapid
in $U-B$ than $B-I$, tracing the evolution of the main-sequence
turn-off.  The reddest clusters in NGC~3310 are on the order of
a few hundred Myr old.

In contrast to the clusters, the diffuse light spans a vary narrow range
of colors.  This agrees well with continuous star formation models.  The
model comparison suggest durations range from 5 to 300 Myr for the
diffuse light, with most of the emission implying $\sim 10$ to 100 Myr
durations.

The overall picture is fairly star formation in both the clusters and
consistent: the diffuse light has been ongoing for several tens or up to
a few hundred Myr.  This is several crossing times of the
starburst ring.

\section{Discussion}

Other researchers have also reported extended star formation
durations within starbursts.  The work of Whitmore \etal\ (1999) on the
spectacular merging ``Antenna'' system (NGC~4038/4039; discussed in
these proceedings by Miller, 2000) reveals clusters with ages up to
$\sim 200$ Myr within the starburst.  This is largely consistent with
the merger timescale of 200 Myr, and a few times the disk crossing
timescale of $\sim 50$ Myr (Barnes 1988).  Calzetti \etal\ (1997; 2000)
report clusters in the blue compact dwarf (BCD) galaxy NGC~5253 with
photometric and spectroscopic ages up to $\sim$ 10 Myr, a few times
larger than the starburst crossing timescale of $\sim 3$ Myr.  Walborn
\&\ Blades (1997) discuss in detail the complex age distribution of
stars and star clusters in the nearest external starburst: 30 Doradus.
Grebel \&\ Chu (2000) expand on a case in point, reporting an age of 25
Myr for Hodge 301, a neighbor of R136a which has an age of 2.5 Myr.  The
projected turbulent velocity dispersion crossing time between these
clusters is $\la 1$ Myr.  Clearly even small starbursts have an extended
history.

Ground based observations of BCDs and H{\sc ii} galaxies yield similar
results.  Their strong emission line spectra indicate that they must
contain a substantial young ionizing population.  However, the colors of
their starbursts (determined after carefully subtracting the underlying
populations, and correcting for extinction), are clearly too red to have
been produced by a young ($\la 10$ Myr) instantaneous burst.  Instead
they are consistent with continuous star formation over $\sim$ 10 to 100
Myr timescales (Telles \&\ Terlevich 1997; Marlowe \etal\ 1999).  This
amounts to at least a few crossing times in most starbursts.

Contrary claims for relatively short burst durations have been made,
particularly with regard to Wolf-Rayet galaxies (Schaerer, Contini, \&\
Kunth, 1999) and BCDs (Mas-Hesse \& Kunth 1999) which have star
formation durations estimated to be $\la 4$ Myr\footnote{Mas-Hesse \&\
Kunth note that in several cases BCD data are consistent with $\sim$ 20
Myr continuous star formation.}.  In these studies timescales are often
spectroscopically constrained by the presence of WR features, which are
washed out for durations $\ga 5$ Myr.  However, the spectroscopy can be
misleading as suggested by Fig.~4.  An observer would usually center the
slit on the brightest source within the galaxy, which is likely to be a
cluster, and preferably a young one.  Such a cluster could dominate the
entire spectrum, the majority of the starburst being outside of the
slit. This would give the impression that the whole starburst is
dominated by a short duration burst.  Cases in point are NGC~5253 (Walsh
\&\ Roy, 1987; Schaerer \etal\ 1997) and NGC~3310 (Pastoriza \etal\
1993), both of which show WR spectra, but only over a fraction of the
starburst.

\begin{figure}
\plotone{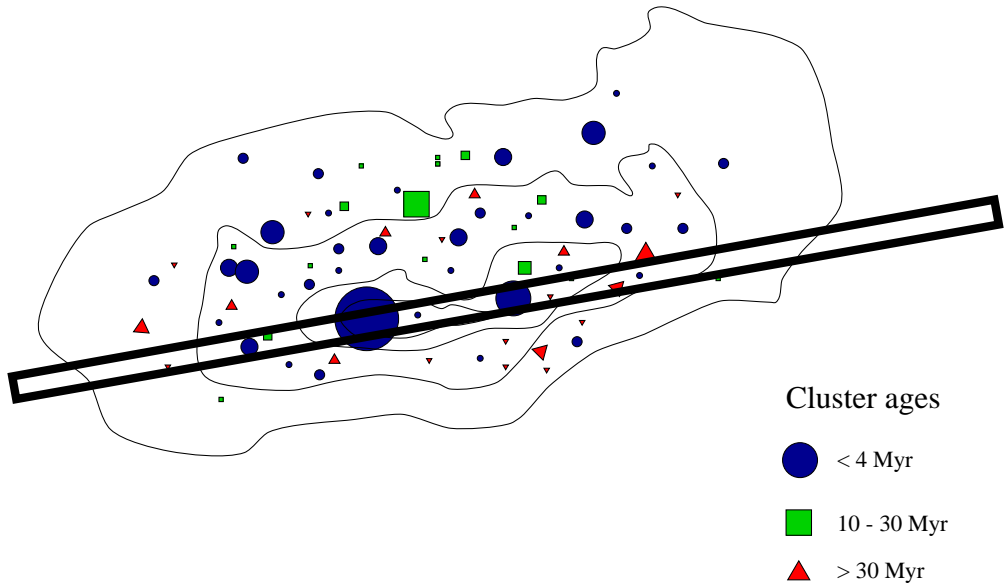}
\caption{Cartoon view of a typical slit orientation (large rectangle) on
a starburst.  The contours represent the distribution of the smooth
light.  The various filled symbols are the clusters with symbol shape
representing age, and size representing brightness.  }
\end{figure}

Another strong voice for short burst durations is Elmegreen (2000) who
claims that star formation occurs over a timescale of only one or two
crossing times, citing various results in the Galaxy and LMC.  The
difference here may in part be due to the velocity used to calculate the
crossing time.  Elmegreen uses the sound velocity whereas I have quoted
measured velocity dispersions of the ISM of the starbursts.  The latter
is supersonic, resulting in lower crossing times, and a
better representation of the true speed that a disturbance travels
within the star forming medium.

If starbursts do last many crossing times there are some important
implications. Firstly, starbursts are sustainable, perhaps even
self-regulating.  The upper-limit to the effective surface brightness of
starbursts reported by Meurer \etal (1997) and Lehnert \&\ Heckman
(1996) is further evidence of this regulation, although it is not clear
what the regulating mechanism is.  Second, burst durations may be longer
than the typical ISM expansion timescales seen in galactic winds (Martin
1998).  Hence, galactic winds are likely to occur in a previously
fractured ISM.  This should increase the efficiency of metal ejection
into the intergalactic medium compared to simulations which usually are
modeled in smooth undisturbed media.

{\bf Acknowledgements}.  I gratefully acknowledge Tim Heckman and Claus
Leitherer, my collaborators on this project.

\parindent=0mm
\bigskip

{\large\bf Comments and discussion}

\bigskip
{\bf Grebel}: The duration of a starburst appears to be determined by the
angular resolution with which one looks at a starburst region.  Consider
for instance 30 Doradus, which can be observed with very high angular
resolution: The massive central cluster R136 has an age of 1-2 Myr;
Hodge 301 has an age of $\sim$ 25 Myr, and there are several other
spatially distinct regions within 30 Dor that have ages between the
above.  The duration of formation in each of these clusters in this
starburst region is only a few Myr.  The entire 30 Dor region is
composed of several smaller ``starbursts'' if they were seen at a
distance of 40 or 100 Mpc.  

{\bf Meurer}: Yes this is a problem.  Even with HST, our method works best on
the nearest systems.

\medskip
{\bf Lan\c{c}on}: It is generally accepted that you worry about internal
extinction in young clusters, not in old clusters.  At what time should
one stop worrying about internal extinction?  Is dust produced by red
supergiant / AGB stars relevant?

{\bf Meurer}: I expect that once a cluster ejects its natal ISM, that it
shouldn't be much affected by internal dust until the starburst region
shuts down.  Hot ($10^{6-7}$ K) gas pervades the entire starburst at a
high filling factor.  Internally produced dust will be quickly swept
away or destroyed.

\medskip
{\bf Burgarella}:  From UV-2000\AA\ observations with the balloon borne FOCA
telescope, we estimated a diffuse light / point like sources to be
closer to 50/50.  This may not be consistent with the ratio that you find
for NGC~3310.  One point to note is that even with a resolution of
$\sim 10$ arcsecs, we probably got smaller star forming regions.  Do you
find that this may be consistent with your view?  We should be careful
about what is diffuse light.

{\bf Meurer}: I suspect that much of the difference is due to resolution.
However much work needs to be done to see the clumpy fraction varies
with properties such as metallicity, potential well depth, and star
formation intensity (surface brightness) The best place to check may be
the Magellanic clouds where UIT images directly resolve the UV
continuum.

\medskip
{\bf Zinneckar}: You mentioned that in NGC~3690 20\%\ of the UV light is in
prominent clusters while 80\%\ is in diffuse UV light. What is the
nature of this diffuse light: scattered light
from the prominent clusters or the superposition of unresolved light
from a more distributed OB star poulation?  What is the spectrum or
color of the diffuse light?

{\bf Meurer}: Calzetti \etal\ (2000) shows that the spectrum of
NGC~5253's diffuse light has narrower {\sc Civ} and Si{\sc iv} lines
than seen in the WR clusters.  Its diffuse light is starting to resolve
into individual stars with HST imaging (e.g.\ M95), so it is unlikely to
be dominated by scattered light.

\medskip
{\bf Schaerer}: Indeed it is important to clarify what kind of
object/region is included in the observations when burst durations are
studied.  Although no exact quantification of this is usually done for
WR ``galaxies'', the short durations mostly found (cf. Schaerer et al.\
1999; Mas-Hesse and Kunth 1999) is likely due to the fact that just one
or a few clusters typically dominate the integrated light in these
objects (most metal-poor BCDs).

\end{document}